\documentclass[12pt]{article}

\usepackage[latin1]{inputenc}
\usepackage{amsmath}
\usepackage{amsthm}
\usepackage{amssymb}
\usepackage{psfrag}
\usepackage{graphicx}
\usepackage{url}

\newtheorem{thm}{Theorem}

\newtheorem{corol}[thm]{Corollary}

\begin{document}

\title{No nonlocal box with uniform outputs is universal}
\author{Fr\'ed\'eric Dupuis$^{\star}$ \quad Nicolas Gisin$^{\dagger}$ \quad Andr\'e Allan M\'ethot$^{\dagger}$\\[0.5cm]
\normalsize\sl $^{\star}$ D\'epartement d'informatique et de recherche op\'erationnelle\\[-0.1cm]
\normalsize\sl Universit\'e de Montr\'eal, C.P.~6128, Succ.\ Centre-Ville\\[-0.1cm]
\normalsize\sl Montr\'eal (QC), H3C 3J7~~\textsc{Canada}\\[-0.1cm]
\normalsize\url{dupuisf}\textsf{@}\url{iro.umontreal.ca}\\[0.5cm]
\normalsize\sl $^{\dagger}$ Group of Applied Physics\\[-0.1cm]
\normalsize\sl Universit\'e de Gen\`eve, rue de l'Ecole-de-M\'edecine 20\\[-0.1cm]
\normalsize\sl 1211 Geneva 4~~\textsc{Switzerland}\\[-0.1cm]
\normalsize\url{{nicolas.gisin,andre.methot}}\textsf{@}\url{physics.unige.ch}
}

\maketitle

\begin{abstract}
We show that standard nonlocal boxes, also known as Popescu-Rohrlich machines, are not sufficient to simulate any nonlocal correlations that do not allow signalling. This was known in the multipartite scenario, but we extend the result to the bipartite case. We then generalize this result further by showing that \emph{no} finite set containing any finite-output-alphabet nonlocal boxes with uniform outputs can be a universal set for nonlocality.
\end{abstract}

\section{Introduction}\label{sec:intro}
\begin{figure}[thb!]
\centering
\psfrag{x}{$x \in \{0,1\}$}
\psfrag{y}{$y \in \{0,1\}$}
\psfrag{a}{$a \in_r \{0,1\}$}
\psfrag{b}{$b \in_r \{0,1\}$}
\psfrag{Alice}{Alice}
\psfrag{Bob}{Bob}
\psfrag{eqn}{$b-a \mod 2 = xy$}
\includegraphics{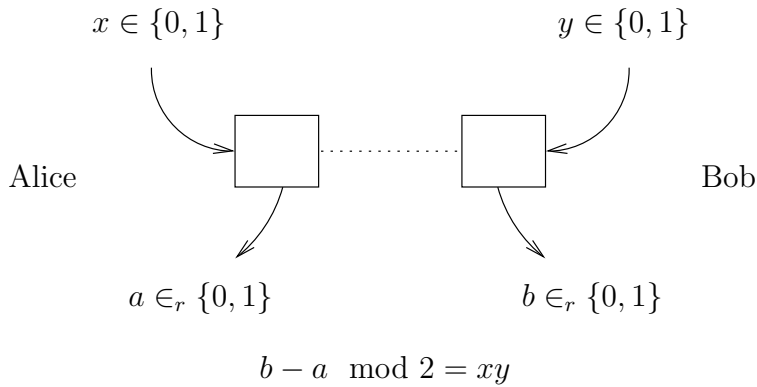}
\caption{The standard nonlocal box. The notation $a \in_r \{0,1\}$
means that $a$ is uniformly distributed over $\{0,1\}$.}
\label{standard-nlb}
\end{figure}

Nonlocality refers to a multi-party process that, while it does not
allow for communication, would classically necessitate communication
for the different parties to perform. One classic example is the
following ``nonlocal box'' (see Figure~\ref{standard-nlb}): imagine
that two parties, henceforth referred to as Alice and Bob, have a
black box into which they can each enter one bit of their choice and
the box gives each of them a random bit such that the exclusive-or
of the output bits is equal to the AND of the input bits~\cite{PR2}.
If one attempts to implement this box without communication in a
classical world, it is easy to show that it is impossible to succeed
more than 75\% of the time~\cite{CHSH}. On the other hand, if the
output bits are always uniformly distributed for all inputs, then it
is clear that this box does not permit communication between Alice
and Bob, since the local probability distribution does not depend on
the parties inputs.

\begin{figure}[thb!]
\centering
\psfrag{x}{$x\in \mathcal{X}$}
\psfrag{y}{$y\in \mathcal{Y}$}
\psfrag{a}{$a\in \mathcal{A}$}
\psfrag{b}{$b\in \mathcal{B}$}
\psfrag{Alice}{Alice}
\psfrag{Bob}{Bob}
\includegraphics{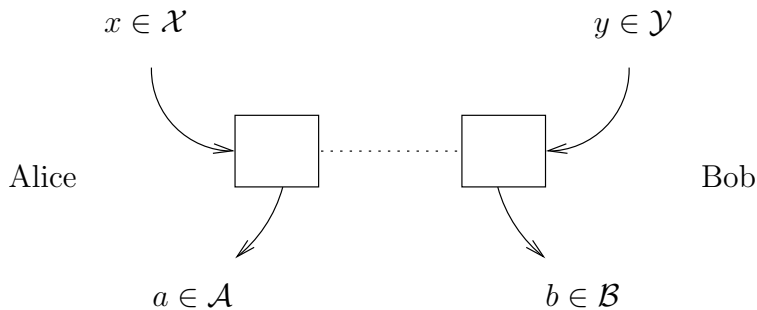}
\caption{A generic nonlocal box. Alice and Bob input $x$ and $y$
respectively, and receive $a$ and $b$ respectively. The resulting
probability distribution $p(a,b|x,y)$ cannot be reproduced
classically without communication, and yet does not itself allow
Alice and Bob to communicate.} \label{generic-nlb}
\end{figure}

We shall call this box the mod2NLB for reasons that will become obvious later. A general nonlocal box is a device such that: given inputs $x$ from Alice and $y$ from Bob, they output values $a$ and $b$ such that the resulting probability distribution $p(a,b|x,y)$ cannot be reproduced classically without communication, yet cannot itself be used to communicate (see Figure~\ref{generic-nlb}).

The original motivation for studying nonlocality is quantum
mechanics. Indeed, in quantum mechanics, entanglement allows us to
achieve nonlocal correlations similar to those described above. John
Bell \cite{bell} was the first to show that measurement on shared
entangled quantum state could produce correlations that cannot be
locally simulated classically. Later, Clauser, Horne, Shimony and
Holt \cite{CHSH} came up with an inequality (the so-called CHSH
inequality) that provided a condition for a certain type of
correlations to be explainable by classical means alone, and showed
that quantum mechanics violated it in some cases. The mod2NLB was
directly inspired by this inequality: the mod2NLB violates the CHSH
inequality to its maximal algebraic value while quantum mechanics
can be used to simulate a mod2NLB with up to approximately 85\%
efficiency~\cite{cirelson80}.

The nonlocality of quantum mechanics has been known for a long time,
but has only recently started to be studied by itself, i.e.\
independently from the study of entanglement. It is hoped that such
an independent study will allow us to understand the implications of
nonlocality in quantum mechanics more thoroughly. Furthermore, there
is proof that entanglement and nonlocality are not the same. The
first example came from~\cite{bgs05}, where it was proved that a
single mod2NLB is not sufficient to simulate a non-maximally
entangled pair of qubits, even though a perfect simulation of all
correlations of the maximally entangled state of two qubits is
possible with only one mod2NLB\cite{cgmp05}. The final proof that
entanglement and nonlocality are different resources came
in~\cite{bm06}, where it was proven that a simulation of $n$
maximally entangled pair of qubits required $\Omega(2^n)$ mod2NLBs.

This asserts that entanglement and nonlocality should be treated as different types of resources. But while we know a fair bit about entanglement, comparatively speaking little is known about nonlocality. For instance, we have been able to isolate the maximally entangled state of two qubits as the ``unit'' of bipartite entanglement, since, together with local operations and classical communication, it allows us to create any other entangled state, provided we have enough copies. Is there an analogous concept for nonlocality? Would it be possible to identify a similar ``unit of nonlocality'', that would allow us to create other bipartite nonlocal correlations? The mod2NLB was the obvious candidate: its minimal size (binary inputs and outputs) and the fact that it violates the CHSH inequality, the only nontrivial inequality at these dimensions, maximally made it very attractive from that point of view.

There are more encouraging signs to support the mod2NLB's claim as the universal resource of nonlocality. One particularly interesting result is that the mod2NLB makes communication complexity~\cite{kn97} trivial~\cite{vandamCC,cleveCC}. That is, if two players are allowed to use mod2NLBs, they can compute any boolean function of their inputs with a single bit of communication, regardless of what the function is.

In light of these facts, it is tempting to think that the mod2NLB
could be considered as a unit of nonlocality that can be used to
generate any other bipartite nonlocal correlation. Some progress has
been made in this direction: in~\cite{BP}, Barrett and Pironio have
shown that mod2NLBs alone can be used to simulate any two-output
bipartite boxes. However, they have shown that there exist
multipartite nonlocal correlations that cannot be simulated by
mod2NLBs alone. What about the bipartite scenario? In~\cite{BLMPPR},
a family of bipartite nonlocal boxes is presented which can generate
every two-input bipartite box. In this paper, we present a
complementary negative result: we show that no \emph{finite} set
containing any general bipartite nonlocal boxes with uniform outputs
can simulate all bipartite nonlocal boxes.

We start, for intuition, by proving the non-universality of the traditional nonlocal box in Section~\ref{sec:mod2nlb}. We prove that a finite number of mod2NLB cannot perfectly simulate the mod3NLB, to be defined at the beginning of Section~\ref{sec:mod2nlb}. In Section~\ref{sec:modpnlb}, we generalize the result by proving that no finite set of finite-output-alphabet nonlocal boxes with uniform outputs can be universal. We then conclude in Section~\ref{sec:conclusion}.

\section{The non-universality of the traditional nonlocal box}\label{sec:mod2nlb}

\begin{figure}[thb!]
\centering
\psfrag{x}{$x \in \{0,1\}$}
\psfrag{y}{$y \in \{0,1\}$}
\psfrag{a}{$a \in_r \{0,1,\dots,p-1\}$}
\psfrag{b}{$b \in_r \{0,1,\dots,p-1\}$}
\psfrag{Alice}{Alice}
\psfrag{Bob}{Bob}
\psfrag{eqn}{$b-a \mod p = xy$}
\includegraphics{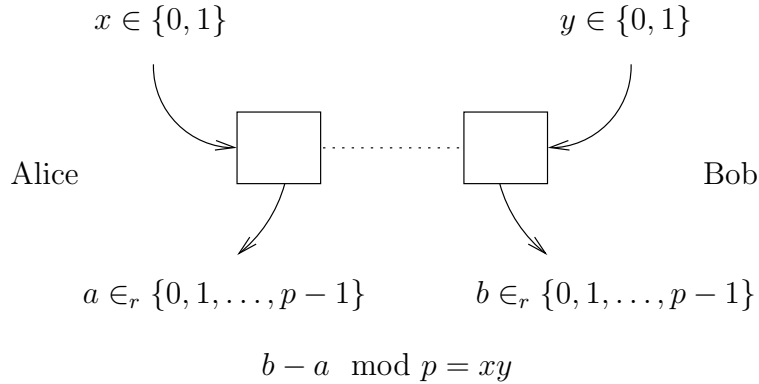}
\caption{A graphical description of the mod$p$NLB.}\label{modpfig}
\end{figure}

We will first begin by introducing the mod3NLB: $x \in \{0,1\}$, $y \in \{0,1\}$, $a \in \{0,1,2\}$, $b \in \{0,1,2\}$, and
\begin{equation}
    \label{boxA3}
    p(a,b|x,y) = \left\{
    \begin{array}{ll}
        \frac{1}{3} & \hbox{ if } b-a = xy \mod 3\\
        0 & \hbox{ otherwise}
    \end{array}
    \right.
\end{equation}
See Figure~\ref{modpfig} for a graphical representation of the general mod$p$NLB. Clearly, this does not allow communication between Alice and Bob, since, taken alone, $a$ is completely independent from $x$ and $y$, and likewise for $b$. The mod3NLB is therefore a valid nonlocal box and a simple extension of the traditional mod2NLB. It would seem reasonable, especially in light of~\cite{BP}, that such a nonlocal box could be simulated by mod2NLBs. However, the following theorem states the opposite.

\begin{thm}\label{thm:mod2NLB}
It is impossible to simulate the mod3NLB exactly using a finite number of mod2NLBs, infinite shared randomness and no communication between the two players.
\end{thm}
\begin{proof}
Let's assume that there exists an algorithm (which may be probabilistic) that can perfectly simulate one instance of the mod3NLB using $N$ mod2NLBs; we will then show that this assumption leads to a contradiction.

First, we can reduce the problem to deterministic algorithms in the following manner: any probabilistic algorithm can be represented as a collection of deterministic algorithms $\alpha_i$, each with a certain probability of being selected. Since we require perfect simulation of the mod3NLB, the outputs of the algorithm must satisfy the equation $b-a = xy$ with probability 1; hence each algorithm $\alpha_i$ with nonzero probability in any probabilistic algorithm must also satisfy this equation with probability 1. For our contradiction, we can therefore restrict ourselves to deterministic algorithms, since a correct probabilistic algorithm exists only if a deterministic algorithm satisfying $b-a = xy$ exists.

Observe first that, for all deterministic algorithms, the output $a$
is completely determined by $x$ and the $N$ output bits that Alice
got from the mod2NLBs, since we can simulate Alice's algorithm using
only those $N+1$ values. Likewise, we can do this on Bob's side to
determine $b$ from $y$ and Bob's mod2NLB outputs. To formalize this,
let $z_A$ be the bit-string that Alice obtained from the mod2NLBs,
and $z_B$ be Bob's bit-string. Then there exist two functions $F_A$
and $F_B$ such that $a = F_A(x, z_A)$ and $b = F_B(y, z_B)$. Note
that $z_A$ and $z_B$ are uniformly distributed on $\{0,1\}^N$. We
can now define the following two probability distributions:
\begin{eqnarray}
p_A(a|x) &=& \Pr\{F_A(x,Z) = a\}\\
p_B(b|y) &=& \Pr\{F_B(y,Z) = b\}
\end{eqnarray}
where $Z$ is a random variable uniformly distributed on $\{0,1\}^N$.

Let us note that $2^N$ is not divisible by 3, therefore $p_A$ and
$p_B$ cannot be uniform for any value of $x$ and $y$. Since we must
be able to simulate the box perfectly, we must at least have:
\begin{eqnarray}
    \label{cdn1} p_A(q|0) &=& p_B(q|0)\\
    \label{cdn2} p_A(q|0) &=& p_B(q|1)\\
    \label{cdn3} p_A(q|1) &=& p_B(q|0)\\
    \label{cdn4} p_A(q|1) &=& p_B(q+1|1)
\end{eqnarray}
where additions are performed mod 3. Condition $(\ref{cdn1})$ comes from the fact that if $x=y=0$, then $b=a$ every time, hence the two marginal distributions must be identical, and therefore $p_A(q|0) = p_B(q|0)$. The other three conditions correspond to similar conditions when the inputs are $(a,b) = (0,1)$, $(a,b) = (1,0)$ and $(a,b) = (1,1)$ respectively.

These conditions lead to a contradiction: $(\ref{cdn1})$ and
$(\ref{cdn3})$ imply that $p_A(q|0) = p_A(q|1)$, which means that
$(\ref{cdn2})$ and $(\ref{cdn4})$ imply that $p_B(q|1) =
p_B(q+1|1)$. Since $p_B(q|1)$ cannot be uniform, we are forced to
conclude that perfect simulation of the mod3NLB with $N$ mod2NLBs is
impossible.
\end{proof}

\section{Generalization to a finite set of nonlocal boxes with uniform outputs}\label{sec:modpnlb}

The result of Section~\ref{sec:mod2nlb} can be generalized to a
finite set of nonlocal boxes, as defined in Section~\ref{sec:intro}
and represented in Figure~\ref{generic-nlb}, where the dimensions of
the output sets are finite, i.e.\ $\vert \mathcal{A}\vert , \vert
\mathcal{B} \vert < \infty$, and the probability distributions
$p(a|x)$ and $p(b|y)$ are both uniform for all $x$ and $y$.
Before turning to the main theorem and its proof, we need to define
the mod$p$NLB in the following manner:
\begin{equation}
    \label{boxAn}
    p(a,b|x,y) = \left\{
    \begin{array}{ll}
        \frac{1}{p} & \hbox{ if } b-a = xy \mod p\\
        0 & \hbox{ otherwise}
    \end{array}
    \right.
\end{equation}
This family of nonlocal boxes was first defined in \cite{BLMPPR}. It was also shown that this family, which is an infinite set, could be used to simulate any two-input bipartite nonlocal box. They also showed that a mod$p$NLB and a mod$q$NLB could simulate a mod$r$NLB, where $r=pq$. Here, we shall prove that for every prime $p$, one instance of a mod$p$NLB cannot be simulated exactly by any combination of a finite number of nonlocal boxes with uniform outputs with $\vert \mathcal{A}\vert , \vert \mathcal{B} \vert < p$. The proof follows that of Theorem~\ref{thm:mod2NLB} very closely.

\begin{thm}\label{thm:modpNLB} For any prime number $p$, the mod$p$NLB cannot be simulated by a finite set of nonlocal boxes with uniform outputs with $\vert \mathcal{A}\vert , \vert \mathcal{B} \vert < p$.
\end{thm}
\begin{proof}
    Let Alice and Bob use a deterministic algorithm to attempt to simulate one instance of the mod$p$NLB, where $p$ is prime, using nonlocal boxes with $\vert \mathcal{A}\vert , \vert \mathcal{B} \vert < p$. Let $M_n$ be the number of NLBs they are using with $|\mathcal{A}| = n$ and $N_n$ be the number of NLBs with $|\mathcal{B}| = n$. Define $z_A$ and $z_B$ to be the strings obtained by Alice and Bob respectively from all their nonlocal boxes; thus $z_A \in \left\{ 0,1 \right\}^{M_2} \times \left\{ 0,1,2 \right\}^{M_3} \times \cdots \times \left\{ 0,1,\dots,p-1 \right\}^{M_{p-1}}$, and $z_B\in \left\{ 0,1 \right\}^{N_2} \times \left\{ 0,1,2 \right\}^{N_3} \times \cdots \times \left\{ 0,1,\dots,p-1 \right\}^{N_{p-1}}$. Again, we also define the functions $F_A$ and $F_B$ such that $a = F_A(x, z_A)$ and $b = F_B(y,z_B)$, and the two probability distributions:
\begin{eqnarray}
p_A(a|x) &=& \Pr\{F_A(x,Z) = a\}\\
p_B(b|y) &=& \Pr\{F_B(y,Z') = b\}
\end{eqnarray}
where $Z$ is a random variable uniformly distributed on $\left\{ 0,1 \right\}^{M_2} \times \left\{ 0,1,2 \right\}^{M_3} \times \cdots \times \left\{ 0,1,\dots,p-1 \right\}^{M_{p-1}}$ and $Z'$ is a random variable uniformly distributed on $\left\{ 0,1 \right\}^{N_2} \times \left\{ 0,1,2 \right\}^{N_3} \times \cdots \times \left\{ 0,1,\dots,p-1 \right\}^{N_{p-1}}$.

Let us note that $2^{M_2} 3^{M_3} \cdots (p-1)^{M_{p-1}}$ has no common divisor with $p$, since $p$ is prime, and thus $p_A(\cdot | x)$ cannot be uniform for any value of $x$. Likewise for $p_B(\cdot | y)$ over $y$. This allows us to get the same contradiction as in the special case of Theorem~\ref{thm:mod2NLB}; in fact, the remainder of the proof is identical. We have therefore shown that no finite set of nonlocal boxes with uniform outputs with $\vert \mathcal{A}\vert , \vert \mathcal{B} \vert < p$ can simulate the mod$p$NLB.
\end{proof}
\begin{corol}
No finite-output-alphabet nonlocal box with uniform outputs is universal.
\end{corol}
\begin{proof}
The proof follows from Theorem~\ref{thm:modpNLB} and the fact that there exists an infinite number of prime numbers. Therefore, for every finite set of nonlocal boxes with uniform outputs, it is possible to define a mod$p$NLB with $p$ a prime number larger than all the output dimensions of the nonlocal boxes in the set.
\end{proof}

\section{Discussion and Conclusion}\label{sec:conclusion}

We have proven that no finite set of finite-output-alphabet nonlocal boxes with uniform outputs can be universal. Therefore only nonlocal boxes with either an infinite number of outputs or nonuniform outputs have a chance for the title. However, there are no compelling candidates in either category, and one might argue that such a box would be even more artificial and less elegant than the traditional mod2NLB.

It is to be noted that our result does not contradict those of~\cite{BP}, since the universality of the family of mod$p$NLBs is defined for binary input nonlocal boxes and requires an infinite set of mod$p$NLBs.

Our result exhibits new difference between nonlocality and
entanglement. The latter has a very simple and attractive universal
resource, the maximally entangled pair of qubits, while the former
has no such things. Our result also suggest that one must be careful
about statements made with the traditional mod2NLB, for it cannot be
associated with a general idea of nonlocality. It is important to
stress at this point that we do not believe research in nonlocal
boxes to be futile. For example, one can still uncover some
intuitions about Nature when studying the mod2NLB.
In~\cite{bblmtu05}, it was proven, using mod2NLBs, that if quantum
mechanics were slightly more nonlocal, it would have drastic and
arguably unbelievable consequences in communication complexity.

This work raises a philosophical question. What is the difference
between entanglement and nonlocality? Why does entanglement have a
universal resource while nonlocality doesn't? It is tempting to
think that it might be related to the fact that we limited the
output dimensions of our nonlocal boxes while measurements on
entangled states can have any number of outputs. However, we would
like to point out that the quantum universality theorem uses quantum
teleportation as its main building block, which requires
measurements with a finite set of possible outputs. We believe that
the answer might be related to the question of the difference
between entanglement measures and nonlocality measures~\cite{ms07}.
In our scenario, we do not allow the participants to use classical
communication, since it is a nonlocal resource. On the other hand,
the universality of the maximally entangled pair of qubits is
established by allowing the participants any resource that do not
increase entanglement: shared randomness, local operations
\emph{and} classical communication. If we take away that last
resource, the universality theorem of entanglement breaks down.
Therefore, nonlocality is directly used in order to generate any
possible entangled state out of a maximally entangled pair of
qubits. What does this entail precisely? We will let the reader
ponder this question.

In a physics lab, generating a general entangled state out of
maximally entangled pairs of qubits will generate some errors and
have some imperfections. One could then wonder if there exists a
nonlocal box or a set of nonlocal boxes that could simulate any
other nonlocal box within a finite error $\epsilon$. We conjecture
that the mod2NLB is sufficient for this, but leave the proof for
future work.

\section*{Acknowledgements}
The authors would like to thank Alain Tapp, Hugue Blier and Gilles Brassard for enlightening discussions on the subject. A.A.M. is especially thankful to Gilles Brassard and the Université de Montréal for the hospitality where this collaboration could be developed.  F. D. is supported by the Natural Sciences and Engineering Research Council of Canada (NSERC) through the Canada Graduate Scholarship program. N.G. and A.A.M. are supported in part by the European Commission under the Integrated Project Qubit Applications (QAP) funded by the IST directorate as Contract Number 015848 and NCCR.

\bibliographystyle{IEEEtran}
\bibliography{IEEEabrv,article}

\end{document}